\documentstyle[12pt]{article}

\newcommand{\bmat}{\left(\begin{array}}
\newcommand{\emat}{\end{array}\right)}
\def\NPB#1#2#3{Nucl. Phys. B{#1} (19#2) #3}
\def\PLB#1#2#3{Phys. Lett. B{#1} (19#2) #3}

\def\PRD#1#2#3{Phys. Rev. D{#1} (19#2) #3}
\def\PRL#1#2#3{Phys. Rev. Lett. {#1} (19#2) #3}

\def\yzero{\smash{\hbox{$y\kern-4pt\raise1pt\hbox{${}^\circ$}$}}}

\def\beq{\begin{equation}}
\def\eeq{\end{equation}}
\def\beqa{\begin{eqnarray}}
\def\eeqa{\end{eqnarray}}

\def\-{\hphantom{-}}
\def\ov{\overline}
\def\s2{\frac{1}{2}}

\def\beq{\begin{equation}}
\def\eeq{\end{equation}}
\def\beqa{\begin{eqnarray}}
\def\eeqa{\end{eqnarray}}
\def\tr{{\rm tr \,}}
\def\Tr{{\rm Tr \,}}
\def\diag{{\rm diag \,}}
\def\IF{\relax{\rm I\kern-.18em F}}
\def\II{\relax{\rm I\kern-.18em I}}
\def\IP{\relax{\rm I\kern-.18em P}}

\def\cp{{\cal P}}
\def\IC{\bf C}
\def\IZ{\bf Z}
\def\IR{\bf R}
\def\IS{\bf S}
\def\z2z2{$\IC^3/(\IZ_2\times\IZ_2)$}

\def\id{{\bf 1}}

\def\NN{{\cal N}}
\def\Dsl{\,\raise.15ex\hbox{/}\mkern-13.5mu D} 

\def\IT{\bf T}

 \def\cp#1{\relax\ifmmode {\IP\kern-2pt{}_{#1}}\else $\IP\kern-2pt{}_{#1}$\=fi}

\begin{document}

\makeatletter \@addtoreset{equation}{section} \makeatother
\renewcommand{\theequation}{\thesection.\arabic{equation}}
\pagestyle{empty}
\vspace*{-.5in}
\rightline{CAB-IB 2918200}
\rightline{CERN-TH/2000-321}
\rightline{CTP-MIT-3041}
\rightline{FTUAM-00/22}
\rightline{IFT-UAM/CSIC-00-36}
\rightline{\tt hep-th/0011073}

\begin{center}
\LARGE{\bf
$D=4$ Chiral String Compactifications from Intersecting Branes
\\[10mm]}
\medskip
\large{G.~Aldazabal$^{1}$, S. Franco$^2$,
L.~E.~Ib\'a\~nez$^3$, R. Rabad\'an$^3$, A. M. Uranga$^4$
\\[2mm]}
\small{$^1$ Instituto Balseiro, CNEA and CONICET,\\[-0.3em]
 Centro At\'omico Bariloche, 8400 S.C. de Bariloche, Argentina.\\[1mm]

$^2$ Center for Theoretical Physics,\\[-0.3em]
Massachusetts Institute of Technology,
Cambridge MA 02139, U.S.A.\\[1mm]
$^3$ Departamento de F\'{\i}sica Te\'orica C-XI
and Instituto de F\'{\i}sica Te\'orica  C-XVI,\\[-0.3em]
Universidad Aut\'onoma de Madrid,
Cantoblanco, 28049 Madrid, Spain.\\[1mm]
$^4$ Theory Division, CERN, CH-1211 Geneva 23, Switzerland.
\\[3mm]}

\smallskip

\small{\bf Abstract} \\[3mm]
\end{center}

\begin{center}
\begin{minipage}[h]{14.5cm}
{\small Intersecting D$p$-branes often give rise to chiral fermions living
on their intersections. We study the construction of four-dimensional
chiral gauge theories by considering configurations of type II
D$(3+n)$-branes wrapped on non-trivial $n$-cycles on $\IT^{2n}\times
(\IR^{2(3-n)}/\IZ_N)$, for $n=1,2,3$. The gauge theories on the four
non-compact dimensions of the brane world-volume are generically chiral
and non-supersymmetric. We analyze consistency conditions (RR tadpole
cancellation) for these models, and their relation to four-dimensional
anomaly cancellation. Cancellation of $U(1)$ gauge anomalies involves a
Green-Schwarz mechanism mediated by RR partners of untwisted and/or
twisted moduli. This class of models is of potential phenomenological
interest, and we construct explicit examples of $SU(3)\times SU(2)\times
U(1)$ three-generation models. The models are non-supersymmetric, but the
string scale may be lowered close to the weak scale so that the standard
hierarchy problem is avoided. We also comment on the presence of scalar
tachyons and possible ways to avoid the associated instabilities. We
discuss the existence of (meta)stable configurations of D-branes on
3-cycles in $(\IT^2)^3$, free of tachyons for certain ranges of the
six-torus moduli.}

\end{minipage}
\end{center}
\newpage
\setcounter{page}{1} \pagestyle{plain}
\renewcommand{\thefootnote}{\arabic{footnote}}
\setcounter{footnote}{0}

\section{Introduction}

D-branes have turned out to be a key ingredient in our present understanding
of the structure of string theory. Interestingly, the fact that D-branes
contain gauge fields localized on their world-volume has also suggested
new scenarios for string phenomenology and phenomenology beyond the
standard model in general (see e.g.\cite{lykken,aadd,otherbw,evenmore,aiqu}).
From this point of view, it is important to explore different
configurations of branes which can lead to interesting features for
phenomenological model building.

An important observation \cite{bdl} is that intersecting D-branes in flat
space may give rise to chiral fermions propagating on the intersection of
their world-volumes, arising from open strings stretching between the
D-branes. Hence, it is natural to consider the construction of
four-dimensional chiral models by compactifications including intersecting
D-branes. In this framework, the compactification space can be essentially
flat, like a six-torus, since chirality arises from fermions at the
intersection of D-brane world-volumes, and does not depend so much on the
holonomy group of the ambient space. This is in contrast with more familiar
compactifications with D-branes, like type IIB orientifolds \cite{orientold,
orientnew,orientfour} or heterotic string compactifications \cite{chsw,ovrut},
where chirality arises due to the ambient space being Calabi-Yau threefold.

In this paper we perform a systematic exploration of configurations of
D$(3+n)$-branes wrapped on $n$-cycles in an $2n$-dimensional torus $\IT^{2n}$,
and sitting at a point in a transverse $6-2n$-dimensional space ${\bf B}$.
We are interested in configurations leading to chiral four-dimensional
gauge theories after reduction on the torus. Chirality is automatically
achieved for D6-branes on 3-cycles on $\IT^6$. However, for models with
D4- or D5-branes, if the point in ${\bf B}$ at which the D-branes sit is
smooth, the resulting intersection lead to vector-like matter. Chiral
matter at the intersections can be obtained by considering branes
sitting at singular points in ${\bf B}$. We will center on abelian
orbifold singularities, whose local geometry can be modeled as
$\IR^{6-2n}/\IZ_N$.

Hence we consider configurations of stacks of D$(3+n)$-branes wrapped on
$n$-cycles in $\IT^{2n}\times \IR^{6-2n}/\IZ_N$. Each stack of D-branes
gives rise to gauge factors, while open strings stretched between them give
rise to chiral fermions propagating on the intersections. The resulting
four-dimensional gauge theories are of potential phenomenological interest.
When the singularity is embedded in a globally compact $(6-2n)$-dimensional
variety, one obtains a full-fledged compactification, where gravity is
also four-dimensional.

Notice that the case of $n=0$ corresponds to configurations of D3-branes
at $\IR^6/\IZ_N$ singularities, which were employed in \cite{aiqu} to
build realistic gauge theories (see \cite{singus} for more formal applications
of these systems).Full-fledged compactifications were subsequently
obtained by embedding the singularities in global compact geometries. Our
approach here is similar in spirit to the bottom-up approach introduced in
\cite{aiqu}, although we mainly center on local features in the present
paper. The models also present a number of interesting new properties.

The opposite extreme case, $n=3$, corresponds to D6-branes wrapped on
3-cycles in $\IT^6$. Configurations of this type have appeared in
\cite{bgkl}, but in the presence of an additional orientifold
projection \footnote{Other models with branes at angles and orientifold
and orbifold projections have appeared in \cite{bgk}.}. This projection is
not a necessary ingredient, and it does not improve the phenomenological or
theoretical features of the model, hence we choose not to include it. In
particular, this allows to get around the orientifold symmetry constraints
in \cite{bgkl}, which prevented the appearance of three-generation models.
Without orientifold action, three-family models are easy to build, and we
present a specific example.

In this paper we perform a detailed analysis of the configurations for
$n=1,2,3$, their construction with explicit examples, and the main features
of the resulting four-dimensional theories. We determine the tadpole
cancellation conditions, their geometrical interpretation, and their
interplay with the cancellation of chiral four-dimensional anomalies.
Interestingly, we find that the theories contain several anomalous $U(1)$'s,
and that their anomalies are cancelled by a generalized Green-Schwarz
mechanism. The fields that mediate this mechanism correspond, for D6-branes
wrapped on 3-cycles on $\IT^6$, to {\em untwisted} closed string modes, in
contrast with the situation in other string constructions. For $n=1,2$
the exchanged fields correspond to reduction on $\IT^{2n}$ of fields in
twisted sectors of $\IR^{6-2n}/\IZ_N$.

The models are generically non-supersymmetric, even if the orbifold twist
is chosen to preserve some bulk supersymmetry. This leads to two important
issues. First, although the discussion of more phenomenological aspects in
these constructions will appear elsewhere \cite{afiru2}, we would like to
mention here the question of scales. Even though models are
non-supersymmetric, it is possible to avoid a hierarchy problem in any
realistic application, by lowering the string scale down to a TeV. This is
possible, i.e. consistent with a large four-dimensional Planck mass, for
models with D4- or D5-branes, in the usual way, by simply taking the
transverse $(6-2n)$-dimensional space ${\bf B}$ large enough. Notice that
the $2n$-torus should remain small (with compactification scale $\approx$
TeV) to avoid too light KK resonances of gauge bosons. Hence, as observed
in \cite{bgkl}, for models with D6-branes solving the hierarchy problem
by large volume compactification is not possible. It is interesting to
consider them, however, in case another mechanism is eventually devised.
On the other hand, let us emphasize again that a low string scale is
consistent with low-energy physics in models with D4- or D5-branes, with a
large transverse volume.

The second comment concerns the generic presence of tachyons at brane
intersections, which signal an instability against recombining intersecting
branes into a single smooth one. Interestingly, for the case of D6-branes
on 3-cycles, there exist brane orientations such that the brane recombination
process is not energetically favoured, since it implies an increase of the
wrapped volume. The corresponding intersection hence does not lead to
tachyonic states. Hence it is in principle possible to construct compact
models of D6-branes on $\IT^6$ where all intersections have this property,
and the resulting model is (meta)stable, as we discuss in some detail.
In models with D4- or D5-branes, it is possible to construct models
where most of the tachyons at intersections are projected out by the
$\IZ_N$ orbifold twist in the quotient singularity.

In any event, we think it is also interesting to consider models with
a small set of tachyons. Recent developments (see \cite{revnbps} for a
review) have suggested that much can be learnt by considering unstable
configurations in string theory and their decay. On the speculative side,
a possible phenomenological application of these ideas would be to interpret
the tachyon condensation process as a Higgs mechanism, in which the two
gauge factors associated to the intersecting branes break to a smaller
subgroup carried by the recombined brane. In fact, it is possible to
construct explicit semirealistic models of D4-branes, where the only
tachyons have the quantum numbers of Standard Model Higgs multiplets. It
is tempting to speculate that the effect of the instability is Higgs
breaking of electroweak symmetry (see \cite{afiru2} for further details).

The paper is organized as follows. In Section 2 we discuss generalities
about the configuration of intersecting branes, and the spectrum arising
on their world-volume and on their intersections. In Section~3 we discuss
the construction of models of D6-branes wrapped on 3-cycles in $\IT^6$. We
analyze their spectrum, the tadpole cancellation conditions and their
interpretation, and cancellation of non-abelian and mixed $U(1)$
anomalies. We also present several explicit examples, e.g. leading to
three-generation Standard Model gauge sectors. We also comment on the
possibility of understanding tachyon condensation as symmetry breaking by
a Higgs mechanism. A similar analysis is carried out for configurations of
D4-branes in Section 4, and of D5-branes in Section~5. Section~6 contains
our final remarks.

\section{Intersecting Dp-branes}

Let us start by considering some generic properties of the spectrum for
branes at angles. We start considering D-branes wrapped on
$d$-cycles $\IT^{2d}$
We choose a factorizable $\IT^{2d}$, product of $d$ two-dimensional
rectangular tori $\IT^2_I$ parameterized by compact coordinates $X_1^I,
X_2 ^I$, with radii $R_1^I, R_2 ^I$, with $I=1,\dots,d$. We introduce $K$
different sets of $N_a$ coincident D$p_a$- branes, labeled by an index
$a$, $a=1\dots,K$. Each set wraps around a 1-cycle $\Pi^I_a$, of type
$(n_a^I, m_a^I)$, on each of the $d$ two-tori. Namely, it wraps $n_a^I$
times around the $X_1^I$ direction and $m_a^I$ times around the $X_2^I$
direction. The angle of these branes with the $X_1^I$ axis is hence given
by
\begin{equation}\label{angle}
\tan\vartheta _a^I= \frac{m_a^I R_2^I}{n_a^I R_1^I}
\end{equation}
with an obvious modification for skewed two-tori.

The compactification preserves all 32 supersymmetries of type II theory in
the closed string sector. The sector of open strings stretching between
Dp$_a$-branes within the same set, preserves 16 supersymmetries, hence
giving rise to the corresponding gauge supermultiplet with gauge group
$U(N_a)$ \footnote{If the wrapping numbers $(n,m)$ are not coprime,
$r=\gcd(n,m)\neq 1$, the D-brane is multiwrapped $r$ times over the cycle
$(n/r,m/r)$. This state can be equivalently described as $r$ D-branes on
$(n/r,m/r)$ with an order $r$ permutation wilson line turned on. For $N$
such multiwrapped branes, the world-volume gauge group is $U(N)^{r}$.
We thank R.~Blumenhagen, B.~K\"ors and D.~L\"ust for discussion on
multiwrapped branes.}.
This piece of the spectrum is non-chiral, so the only source of chiral
fields is the sector of open strings stretched between different sets of
branes.

The spectrum of such sectors has been studied in \cite{bdl}. World-sheet
bosonic fields for open strings stretching between D$p_a$- and
Dp$_b$-branes, at a relative angle $\vartheta_{ba}^I= (\vartheta_b^I-
\vartheta _a^I)/\pi$, (given in `units of $\pi$' for convenience), in
the $I^{th}$ two-torus, satisfy the boundary conditions
\begin{eqnarray}
\label{bcab}
\sin \vartheta_a^I \partial _{\sigma}X_1^{I}- \cos\vartheta_a^I \partial
_{\sigma}X_2^I & =& 0 \nonumber\\
\sin \vartheta _a^I \partial _{t}X_2^{I}-\cos\vartheta_a^I \partial
_{t}X_1^I &=& 0
\end{eqnarray}
at $\sigma=0$, and a similar equation for $\sigma= \pi$ with $a\rightarrow
b$. Corresponding equations are satisfied by fermionic coordinates. Such
boundary conditions lead to {\it twisted} mode expansions, with twist
given by the relative angle $\vartheta_{ba}^I$ between branes. For
instance, one obtains worldsheet fermionic modes $\psi^{I}_{r_-}$,
$\psi^{I}_{r_+}$, with modes $r_{\pm} =n\pm \vartheta_{ba}^I + \nu$, where
$n$ is integer and $\nu =0,1/2$ for R and NS boundary conditions
respectively. No windings or KK momenta are allowed for non trivial angles.
Antiparticles of states in the $ab$ sector appear in the $ba$ sector.

We are mainly interested in four-dimensional intersections, hence we consider
the cases of D$(3+n)$-branes wrapped on $n$-cycles on $\IT^{2n}$, for
$n=1,2,3$. As mentioned in the introduction, configurations $n<3$ would
lead to non-chiral intersections, hence we will eventually turn to
configurations with singular transverse spaces, namely D$(3+n)$-branes on
$n$-cycles in $\IT^{2n}\times \IR^{6-2n}/\IZ_N$. Before that, it is
convenient to discuss the simpler case of $\IT^{2n}\times \IR^{6-2n}$ in
this section, namely D6-branes on $\IT^6$, D5-branes on $\IT^4\times \IC$,
D4-branes on $\IT^2 \times  \IC^2$.

The mass operator for strings stretching between branes in the $a^{th}$
and $b^{th}$ set, which make an angle $\vartheta ^I \equiv \vartheta_{ab}^I$
on the $I^{th}$ two-torus is \cite{arjab,bdl}

\begin{equation}
\label{massop}
\alpha' M_{ab} ^2=\frac{Y^2}{4\pi^2\alpha'}+ N_{\nu}+\nu \sum _{I=1}^d\vartheta_{ab}^I -\nu
\end{equation}

where $Y^2$ measures the length of the stretched string (minimal distance
between branes for minimum winding states), and $N_{\nu} $ is the
number operator given by

\begin{eqnarray}
N_0 & =&  \sum_{n>0}(\alpha_{-n}.\alpha_{n}+
  r \psi_{-r}\psi_r) + \sum _{I=1}^d [\sum_{n>0}(\alpha_{-n_+}^I.\alpha_{n_+}^I+\alpha_{-n_-}^I.\alpha_{n_-} ^I) + \alpha_{-\vartheta ^I }^I\alpha_{\vartheta ^I}^I] \nonumber \\ 
 &+ &  \sum _{I=1}^d [\sum_{n>0}(r_-  \psi_{-r_+}^I.\psi_{r_+}^I+
r_- \psi_{-r_-}^I.\psi_{r_-}^I) + \vartheta ^I  \psi_{-\vartheta ^I}^I\cdot
\psi_{\vartheta ^I }^I]
\end{eqnarray}
for the R sector ($\nu=0$) and by
\begin{eqnarray}
N_{\frac12} & =&  \sum_{n>0}(\alpha_{-n}.\alpha_{n}+  r \psi_{-r}\psi_r )+ 
 \sum _{I=1}^d  [\sum_{n>0}( \alpha_{-n_+}^I.\alpha_{n_+}^I+\alpha_{-n_-}^I.\alpha_{n_-} ^I) + \alpha_{-\vartheta ^I }^I\alpha_{\vartheta ^I}^I]  \nonumber\\
& &  +
 \sum _{I=1}^d  \sum_{n=0}(r_+ \psi_{-r_+}^I.\psi_{r_+}^I+
r_- \psi_{-r_-}^I.\psi_{r_-}^I)  \nonumber\\
\end{eqnarray}
in the NS($\nu=\frac12$) sector.

 The $\vartheta_{ab}^I$ in the mass equation (\ref{massop}) arises from normal
ordering of twisted zero modes and it cancels out in the R sector.
 In the derivation of this  expression we
have assumed that $0 \leq \vartheta_{ab}^I \leq\frac{1}2$,
so oscillators modes as above are correctly normal ordered. For negative
angles one should replace $\vartheta_I \rightarrow |\vartheta_I|$.

The spectrum can be described in bosonic language as follows. We introduce
a four-dimensional twist vector $v_{\vartheta}$, whose $I^{th}$ entry is
given by $\vartheta_{ab}^I$. The GSO projected states are labeled by a
four-dimensional vector $r+v_{\vartheta}$, where $r_I\in \IZ,\IZ+\frac 12$
for NS, R sectors respectively, and $\sum_I r_I={\rm odd}$. The last entry
provides four-dimensional Lorentz quantum numbers. The mass of the states
is then given by
\beq
\alpha' M_{ab}^2 = \frac{Y^2}{4\pi^2\alpha'}+
N_{bos}(\vartheta)+ \frac{(r+v)^2}{2} -\frac 12 + E_{ab}
\eeq
with
\beqa
E_{ab}= \sum_I \frac 12 |\vartheta_I|(1-|\vartheta_I|)
\eeqa
and $N_{bos}(\vartheta)$ is a contribution from bosonic oscillators.

Let us discuss the computation of lowest lying states in the different
models. As mentioned above, models for $n<3$ have a non-chiral spectrum,
as is easily seen from the fact that all massless states can be
made massive in a continuous way by increasing the separation $Y^2$ in
transverse space.

We first consider the case of D4-branes on $\IT^2\times \IC^2$. In the NS
sector, the lowest mass state allowed by GSO projection \cite{bdl}
corresponds to $\psi_{(\vartheta^1 -1/2)}^I|0>_{NS}$, or
$r+v=(-1+\vartheta^1,0,0,0)$ in bosonic language. Its mass is given by
\begin{equation}
\alpha^{\prime} M_1^2= \frac{Y}{4\pi ^2 \alpha^{\prime}}-
\frac 12 |\vartheta^1|
\end{equation}
Thus, a tachyon is generated when D4 branes come closer than the critical
distance $Y^2=2\pi^2\alpha'|\vartheta^1|$. The tachyon signals an instability
against joining the intersecting branes into a single one, which then
wraps a one-cycle in the homology class $[\Pi_a]+[\Pi_b]$, namely a
$(n_a+n_b,m_a+m_b)$ cycle on $\IT^2$. In Section 4 we discuss how to use a
$\IZ_N$ orbifold twist to project out some of these tachyons.

The R groundstate contains four fermions that become massless at zero
transverse distance. They are given by
\begin{eqnarray}
\begin{array}{ccc}
(-\s2+\vartheta^1, -\s2, -\s2,+\s2)  \quad & ; & \quad
(-\s2+\vartheta^1, -\s2, +\s2,-\s2) \\
(-\s2+\vartheta^1, +\s2, -\s2,-\s2)  \quad & ; & \quad
(-\s2+\vartheta^1, +\s2, +\s2,+\s2)
\end{array}
\end{eqnarray}
There are two pairs of opposite chirality spinors, so the spectrum is
non-chiral. A possibility to obtain a chiral spectrum is to project out
some of the above fermions, for instance by locating the D4-branes at
$\IC^2/\IZ_N$ singularities in transverse space, see Section 4.

\medskip

In the case of configurations of D5-branes on $\IT^2 \times \IT^2 \times
\IC$, open strings at intersections have a twist vector $(\vartheta_1,
\vartheta_2,0,0)$. In the NS sector, assuming $0<\vartheta_I<1$ the lowest
mass NS states correspond to $\psi_{-1/2+\vartheta_1}|0>$,
$\psi_{-1/2+\vartheta_2}|0>$, or in bosonic language
$(-1+\vartheta_1,0,0,0)$, $(0,-1+\vartheta_2,0,0)$. Their masses
are $M^2_1=\frac1{2}(\vartheta_2-\vartheta_1)$ and $M^2_2=-\frac1{2}
(\vartheta_2-\vartheta_1)=-M^2_1$, respectively.
Thus, unless $|\vartheta_2|=|\vartheta_1|$, in which case the intersection
preserves some supersymmetry, there is always a tachyonic state. The R
spectrum contains a set of non-chiral massless fermions, corresponding to
the states $(-\s2+\vartheta^1,-\s2+\vartheta^2,-\s2,\s2)$ and
$(-\s2+\vartheta^1,-\s2+\vartheta^2,\s2,-\s2)$. Again, tachyon elimination
and chirality may be obtained by imposing an orbifold projection, namely
by considering D5-branes wrapped on $\IT^4$ and located at the origin of a
$\IC/\IZ_N$ singularity, as we do in section 5.

As mentioned, a chiral spectrum is obtained for D6-brane intersections on
$\IT^6$. The twist vector is now given by $(\vartheta_1,\vartheta_2,
\vartheta_3,0)$. In the NS sector, the lowest lying states, for
$0\leq\vartheta^I\leq1$, are given by
$(-1+\vartheta^1,\vartheta^2,\vartheta^3,0)$,
$(\vartheta^1,-1+\vartheta^2,\vartheta^3,0)$,
$(\vartheta^1,\vartheta^2,-1+\vartheta^3,0)$, and
$(-1+\vartheta^1,-1+\vartheta^2,-1+\vartheta^3,0)$. As discussed in more
detail in Section~3, some of them may be tachyonic, but not necessarily.
In the R sector, we obtain a single chiral fermion, given by
$(-\s2+\vartheta^1, -\s2+\vartheta_2, -\s2+\vartheta_3,+\s2)$.

We conclude by emphasizing an important point. Branes wrapped on
cycles generically intersect at multiple points, hence the above states
in mixed $ab$ sectors appear in several copies, this multiplicity being
given by the intersection number of the corresponding wrapped cycles.
(If e.g. one of the branes, say the $b^{th}$ has non-coprime $(n,m)$,
the multiplets in the $ab$ sector transform as $\sum_{l=1}^r
{\tilde I}_{ab}(N_a,N_{b,r})$ under the gauge group $U(N_a)\times
U(N_b)^r$, and ${\tilde I}_{ab}=I_{ab}/r$).

\section{D6-branes wrapping at angles on $(\IT^2)^3$}

\subsection{Construction}

In this section we consider type IIA theory compactified on a
factorizable $\IT^6$. We consider a configuration containing $K$ stacks of
$N_a$ D6-branes, $a=1,\ldots,K$, wrapped on three-cycles $\Pi_a$ obtained
as the product of one-cycles $(n_a^I,m_a^I)$ on each of the three two-tori
$I=1,2,3$ \footnote{This type of configuration is a particular case of
configurations of D6-branes wrapped on special lagrangian cycles in a
Calabi-Yau threefold (see \cite{kachru} for recent discussions).}. In
\cite{bgkl} this kind of D6-brane configurations were considered in the
presence of an orientifold projection. Since the projection is not required
for consistency, we prefer not to impose this restriction and keep our
analysis general.

The models admit a T-dual description \cite{bdl,bgkl} in terms of type IIB
compactified on a T-dual torus ${\tilde {\IT}}^6$ (with the Kahler and
complex structure on each two-tori exchanged with respect to the original
one), with a set of D9-branes (and anti-D9-branes), with wrapping numbers
$n_a^I$ and world-volume magnetic flux with charge $m_a^I$ along the
$I^{th}$ two-torus. (Models with such fluxes and orbifold and orientifold
projections have appeared in \cite{aads}). Even though we phrase our
discussion in D6-brane language, we will find it useful to  occasionally
turn to this T-dual picture.

The configuration can be described by a free world-sheet CFT, and the
consistency conditions (tadpole cancellation conditions) can be analyzed
by usual factorization of one-loop amplitudes. They read
\beqa
\begin{array}{lcl}
\sum_a N_a n_a^1 n_a^2 n_a^3 = 0 & \quad &
\sum_a N_a n_a^1 m_a^2 m_a^3 = 0 \\
\sum_a N_a m_a^1 n_a^2 n_a^3 = 0 & \quad &
\sum_a N_a m_a^1 n_a^2 m_a^3 = 0 \\
\sum_a N_a n_a^1 m_a^2 n_a^3 = 0 & \quad &
\sum_a N_a m_a^1 m_a^2 n_a^3 = 0 \\
\sum_a N_a n_a^1 n_a^2 m_a^3 = 0 & \quad &
\sum_a N_a m_a^1 m_a^2 m_a^3 = 0
\end{array}
\label{tadpole}
\eeqa
In the D6-brane picture, they are equivalent to the condition that the
homology classes $[\Pi_a]$ of the cycles $\Pi_a$ wrapped by the D6-branes,
counted with multiplicity $N_a$, add up to zero. Denoting by $[a_I]$,
$[b_I]$ the homology classes of the $(1,0)$ and $(0,1)$ basis cycles in
the $I^{th}$ two-torus, we have
\beqa
[\Pi_a] & = & (n_a^1\, [a_1] + m_a^1\, [b_1])\otimes (n_a^2\, [a_2]
+ m_a^2\, [b_2]) \otimes (n_a^3\, [a_3] + m_a^3\, [b_3])
\label{homology}
\eeqa
and (\ref{tadpole}) can be recast as
\beqa
\sum_a N_a [\Pi_a] = 0
\label{zeroclass}
\eeqa
The vanishing of the total homology class is required by consistency with
the equations of motion for the RR 7-form, under which the D6-branes are
electrically charged
\beqa
d*H_8 & = & \sum_a N_a \delta(\Pi_a)
\eeqa
where $H_8$ is the field strength of the 7-form, and $\delta(\Pi_a)$ is
a three-form supported at the location of the D6$_a$-branes, the Poincare
dual of $[\Pi_a]$. Since $d*H_8$ is exact, the above equation in homology
becomes (\ref{zeroclass}).

In the language of D9-branes with fluxes, conditions (\ref{tadpole})
receive the following interpretation. In the presence of background
magnetic fluxes, D9-branes carry charges under RR forms of all even
degrees, due to the WZ world-volume couplings \cite{lidouglas}. The above
tadpole conditions amount to the cancellation of overall D9-, D7$_I$-,
D5$_I$- and D3-brane charges, (where D5$_I$- and D7$_I$-branes, are
wrapped on, or transverse to, the $I^{th}$ two-torus, respectively). This
is required for consistency of the equations of motion of the corresponding
RR forms, i.e. the T-dual statement to our argument in the D6-brane
picture.

\medskip

 From our discussion in Section 2, the four-dimensional field theory
arising after compactification of the D6-branes on the torus contains
chiral fermions arising from brane intersections, hence a priori have
phenomenological interest. They are also non-supersymmetric, but in principle
the existence of tachyon-free stable configuration is not excluded, see
section 3.4.

Let us obtain the massless (and tachyonic) four-dimensional spectrum. The
$6_a6_a$ sector has unbroken $\NN=4$ supersymmetry, and leads, in
component fields, to $U(N_a)$ gauge bosons, six real scalars in the
adjoint representation and four Majorana fermions in the adjoint as well.
In the mixed $6_a6_b$ and $6_b6_a$ sectors, the field content appears in
general in several replicas \footnote{In the T-dual picture in terms of
D9-branes with magnetic fluxes, the multiplicities arise from the Landau
level multiplicities.}, due to the multiple intersection number
$I_{ab}$ of the cycles $\Pi_a$ and $\Pi_b$, given by
\beqa
I_{ab}=[\Pi_a]\cdot [\Pi_b] =\prod_i(n_a^i m_b^i-m_a^i n_b^i)
\label{defintersec}
\eeqa
In the R sector, we obtain $I_{ab}$ chiral left-handed fermions in the
bifundamental representation $(N_a,{\ov N}_b)$, with the understanding
that a negative multiplicity corresponds to a positive multiplicity of
right-handed fermions. In the NS sector, we obtain a set of $I_{ab}$
bifundamental scalars, whose masses are controlled by the angles
$\vartheta_I$ between the D6$_a$ and the D6$_b$-branes, which depend on
the six-torus moduli. Their masses are given by (assuming $0\leq
\vartheta_i\leq 1$)
{\small \beqa
\begin{array}{cc}
{\rm \bf State} \quad & \quad {\bf Mass} \\
(-1+\vartheta_1,\vartheta_2,\vartheta_3,0) & \alpha' M^2 =
\frac 12(-\vartheta_1+\vartheta_2+\vartheta_3) \\
(\vartheta_1,-1+\vartheta_2,\vartheta_3,0) & \alpha' M^2 =
\frac 12(\vartheta_1-\vartheta_2+\vartheta_3) \\
(\vartheta_1,\vartheta_2,-1+\vartheta_3,0) & \alpha' M^2 =
\frac 12(\vartheta_1+\vartheta_2-\vartheta_3) \\
(-1+\vartheta_1,-1+\vartheta_2,-1+\vartheta_3,0) & \alpha' M^2
= 1-\frac 12(\vartheta_1+\vartheta_2+\vartheta_3)
\label{tachdsix}
\end{array}
\eeqa}
Hence certain intersections may lead to the appearance of tachyons. If
present, they signal an instability against joining the intersecting
branes into a single smooth one. As observed in \cite{angleth}, tachyon
modes arise precisely in the range of $\vartheta_I$'s for which the
joining process is energetically favoured, namely decreases the 3-cycle
volume. In Section 3.4 we discuss the construction of models which, for a
range of six-torus moduli, do not contain tachyons at brane intersections.
Hence the corresponding configurations are protected against recombination
by a energy barrier.

In next section we center of robust aspects of the theory, such us the
chiral fermion content, and potential gauge anomalies. Hence recall
that the gauge group and chiral fermions in the models are
\footnote{In fact, the chiral piece of the spectrum of a set of D6-branes
wrapped on 3-cycles in a threefold (not necessarily Calabi-Yau) has this
form, and  our arguments about cancellation of four-dimensional anomalies
are valid (with some obvious modifications) in this general case.}
\beqa
& \prod_{a=1}^K U(N_a) & \nonumber \\
&\sum_{a<b}\, I_{ab}\, (N_a,{\ov N}_b)&
\label{specsix}
\eeqa
The spectrum is generically chiral, leading to an interesting set of
four-dimensional field theories.

\subsection{Anomaly cancellation}

\subsubsection{Non-abelian anomalies}

Following \cite{bdl,ghm}, the gauge anomaly induced by the chiral fermions
living on each intersection is cancelled by an anomaly inflow mechanism
associated to the intersecting branes (see \cite{inflowcomp} for string
computations of the relevant couplings). Namely, the violation of charge
induced by the anomaly is compensated by a charge inflow from the bulk of
the intersecting branes. This explanation is sufficient in situations
where the branes are infinitely extended. In the compact context, however,
within a single brane the charge `inflowing' into an intersection must be
compensated by charge `outflowing' from other intersections \footnote{This
anomaly flow picture is analogous to that in \cite{hz}.}. Consistency
of anomaly inflow in a compact manifold imposes global constraints on the
configuration.

From the point of view of the compactified four-dimensional effective
field theory, which does not resolve the localization of the different
chiral fermions, these global constraints correspond to cancellation of
triangle gauge anomalies in the usual sense. In fact, the cancellation of
cubic non-abelian anomalies for the gauge factor $SU(N_a)$ in
(\ref{specsix}) reads
\beqa
\sum_{b=1}^K\, I_{ab}\, N_b = 0
\label{anomdsix}
\eeqa
 From the ten-dimensional viewpoint, (\ref{anomdsix}) expresses the
cancellation of inflows from different intersections in the D6$_a$-branes.

By replacing (\ref{defintersec}) in (\ref{anomdsix}), one can see that
tadpole cancellation conditions imply the cancellation of cubic non-abelian
anomalies. Thus, as usual, string theory consistency conditions imply
consistency of the low-energy effective theory. However, tadpole
cancellation conditions are in general much stronger than anomaly
cancellation conditions (see also \cite{uprobe}), a feature also found in
the context of standard type IIB orientifolds \cite{abiu} (see also
\cite{morales,uprobe}).

\subsubsection{Mixed $U(1)$ anomaly cancellation}

Let us turn to mixed $U(1)$ anomalies. Again, anomalies at each intersections
are cancelled by the inflow mechanism \cite{bdl,ghm}. However, the global
consistency of the inflow, or equivalently, cancellation of anomalies from
the perspective of the compactified four-dimensional theory, is in this
case more intricate, and involves a Green-Schwarz mechanism \footnote{The
interplay between the inflow and Green-Schwarz anomaly cancellation
mechanisms has been studied in \cite{iztown} in a different context.}.
Using the fermion spectrum in (\ref{specsix}), the mixed $U(1)_a-SU(N_b)$
triangle anomaly reads
\beqa
{\cal{A}}_{ab} =
\frac 12\, \delta_{ab} \sum_c N_c\, I_{bc} + \frac 12\, N_b\, I_{ab}
\label{mixedsix}
\eeqa
The first piece is proportional to the non-abelian anomaly, and vanishes,
while the last piece is generically non-vanishing even after imposing
tadpole conditions.

We now show that the residual anomaly is cancelled by a generalized
Green-Schwarz mechanism mediated by RR partners of closed string {\em
untwisted} geometric moduli. This situation contrasts with that in type
IIB orientifolds, where $U(1)$ anomalies are cancelled through exchange of
closed string {\em twisted} moduli \cite{iru} (see \cite{sagnan} for the
six-dimensional case, and e.g.\cite{abd,scruse} for subsequent work). It
also differs from that in heterotic compactifications, in not involving
the dilaton multiplet, and in allowing the existence of several anomalous
$U(1)$'s.

Let us consider a D6$_a$-brane wrapped on a 3-cycle $[\Pi_a]$. It has
several relevant world-volume couplings \cite{lidouglas} to the RR 3-form
$C_3$ and its ten-dimensional Hodge dual, the 5-form $C_5$
\beqa
\int_{D6_a} C_3\wedge F_a \wedge F_a \quad ; \quad
\int_{D6_a} C_5 \wedge F_a
\label{couplsix}
\eeqa
In order to obtain the couplings after Kaluza-Klein reduction to four
dimensions, it is convenient to introduce two basis of homology 3-cycles,
$\{ [\Sigma_i] \}$, $\{ [\Lambda_i] \}$, dual to each other, namely
$[\Lambda_i]\cdot [\Sigma_j]  =  \delta_{ij}$. On these two basis, the
D6$_a$-brane 3-cycle $[\Pi_a]$ has the expansions
\beqa
[ \Pi_a ] = \sum_{i} r_{ai} [ \Sigma_i ] \quad ; \quad
[ \Pi_a ] = \sum_{i} p_{ai} [ \Lambda_i ]
\eeqa
Defining the untwisted RR fields $\Phi_i = \int_{[\Lambda_i]} C_3 \quad ;
\quad B_2^i = \int_{[\Sigma_i]} C_5$,
which are Hodge duals in the four-dimensional sense, the couplings
(\ref{couplsix}) read
\beqa
\sum_i p_{ai} \int_{M_4} \Phi_i\, F_a \wedge F_a\quad ; \quad
N_a \sum_{i} r_{ai} \int_{M_4} B_2^i \wedge F_a
\eeqa
where the prefactor $N_a$ arises from normalization of the $U(1)$
generator, as in \cite{iru}. These couplings can be combined in a  GS
diagram where $U(1)_a$ couples to the $i^{th}$ untwisted field, which
then couples to $F_b^2$. The coefficient of this amplitude is (modulo an
$a$, $b$ independent numerical factor)
\beqa
N_a \sum_i\, r_{ai}\, p_{bi} = N_a\, \sum_{i,j}\, r_{ai}\, p_{bj}\,
[\Sigma_i]\cdot [\Lambda_j] = N_a\, [\Pi_a]\cdot [\Pi_b] = N_a I_{ab}
\eeqa
precisely of the form required to cancel the residual
$U(1)_a$-$SU(N_b)^2$ anomaly in (\ref{mixedsix}).

\medskip

The same mechanism may be described in the T-dual picture of D9-branes
with magnetic fluxes. The couplings on the world-volume of D9-branes to
bulk RR fields \footnote{The role of these couplings in anomaly
cancellation in a different class of models has been suggested in
\cite{aads}.} are of the form (wedge products implied)
\beqa
\begin{array}{ccccc}
\int_{D9_a} C_0 \, F_a^5 & ; & \int_{D9_a} C_2 \, F_a^4 & ; & \int_{D9_a}
C_4\, F_a^3 \\
\int_{D9_a} C_6 \, F_a^2 & ; & \int_{D9_a} C_8\, F_a & ; & \int_{D9_a}
C_{10}
\end{array}
\eeqa
In order to obtain the four-dimensional version of these couplings, we
define
\beqa
\begin{array}{cclccl}
C_2^{I} & = & \int_{(\IT^2)_I} C_4 & \quad ; \quad
C_0^{I} & = & \int_{(\IT^2)_I} C_2 \nonumber \\
B_2^{I} & = & \int_{(\IT^2)_J\times (\IT^2)_K} C_6 \quad
 & \quad ; \quad
B_0^{I} & = & \int_{(\IT^2)_J\times (\IT^2)_K} C_4 \quad\nonumber\\
B_2 & = & \int_{(\IT^2)_1\times (\IT^2)_2\times (\IT^2)_2} C_8 & \quad ;
\quad
B_0 & = & \int_{(\IT^2)_1\times (\IT^2)_2\times (\IT^2)_3} C_6
\end{array}
\eeqa
where $I\neq J\neq K\neq I$ in second row.
The fields $C_2$ and $C_6$, and also $C_0$ and $C_8$ are Hodge duals,
while $C_4$ is self-dual. In four dimensions, the duality relations are
\beqa
\begin{array}{cccccc}
dC_0 & = & * d B_2 & \quad ; \quad
dB_0^I & = & * d C_2^I \nonumber \\
dC_0^I & = & - * d B_2^I & \quad ; \quad
dB_0 & = & - * dC_2
\end{array}
\label{hodge}
\eeqa

In the dimensional reduction, one should take into account that
integration of $F_a$ along the $I^{th}$ two-torus yields a factor $m_a^I$.
Also, integrating the pullback of the RR forms on the (multiply wrapped)
D9$_a$-brane over the $I^{th}$ two-torus yields a factor $n^I_a$. We
obtain the couplings
\beqa
\begin{array}{ccc}
N_a\, m^1_a\, m^2_a\, m^3_a \int_{M_4} C_2 \wedge F_a & \quad ; \quad
& n^1_b\, n^2_b\, n^3_b \int_{M_4} B_0 \wedge F_b \wedge F_b \nonumber \\
N_a\, n^I_a\, m^J_a\, m^K_a \int_{M_4} C_2^I \wedge F_a & \quad ; \quad
& n^J_b\, n^K_b\, m^I_b \int_{M_4} B_0^I \wedge F_b\wedge F_b \nonumber \\
N_a\, n^J_a\, n^K_a\, m^I_a \int_{M_4} B_2^I \wedge F_a & \quad ; \quad
& n^I_b\, m^J_b\, m^K_b \int_{M_4} C_0^I \wedge F_b\wedge F_b  \nonumber\\
N_a\, n^1_a\, n^2_a\, n^3_a \int_{M_4} B_2 \wedge F_a & \quad ; \quad
& m^1_b\, m^2_b\, m^3_b \int_{M_4} C_0 \wedge F_b\wedge F_b
\end{array}
\eeqa
As usual, the $N_a$ prefactors arise from $U(1)_a$ normalization.

The GS amplitude where $U(1)_a$ couples to one untwisted field which
propagates and couples to two $SU(N_b)$ gauge bosons is proportional to
\beqa
& & -N_a\,\; m^1_a m^2_a m^3_a n^1_b n^2_b n^3_b +N_a\, \sum_I  n^I_a m^J_a
m^K_a n^J_b n^K_b m^I_b - N_a\, \sum_I n^I_a n^J_a m^K_a n^K_b m^I_b m^J_b +\nonumber \\ & &
N_a\,n^1_a n^2_a m^3_a m^1_b m^2_b m^3_b \;
 = N_a \prod_{I} (n_a^I m_b^I - m_a^I n_b^I) = N_a I_{ab}
\eeqa
as required to cancel the residual mixed $U(1)$ anomaly in (\ref{mixedsix}).

\smallskip

Finally, it is straightforward to check that these theories do not produce
mixed $U(1)$ gravitational anomalies.

Due to the linear couplings between the $U(1)$'s and the closed string
moduli, anomalous $U(1)$'s become massive with a mass of the order of the
string scale. Therefore it is important, for any (phenomenological or not)
application of these models, to determine the precise linear combinations
becoming massive and those staying massless.

One can advance that since there are eight fields mediating the anomaly
cancellation, at most eight $U(1)$ linear combinations can gain mass.
Denoting $Q_a$ the generator of the $a^{th}$ $U(1)$, and writing a general
linear combination as
\beqa
Q = \sum_{a=1}^K \frac{c_a}{N_a} Q_a
\label{lincomb}
\eeqa
non-anomalous $U(1)$s correspond to zero modes of the intersection matrix
$\sum_a c_a I_{ab} = 0$.

We conclude with a brief discussion of Fayet-Iliopoulos terms for the
anomalous $U(1)$'s. An important observation is that the standard
low-energy field theory arguments relating GS mechanism with FI terms in
\cite{dsw} are based on supersymmetry, hence do not directly apply to our
models. Notice however that the string theory diagram giving rise to
linear couplings between anomalous $U(1)$ and closed string modes is a
disk, with boundary on the relevant D6-brane and a closed string mode
insertion. This diagram does not notice the breaking of supersymmetries by
other branes, hence yields superpartner interactions, and in particular a
FI term, proportional to the NS-NS part of untwisted moduli. As opposed to
supersymmetric cases, where the FI terms are not renormalized, in the
present non-supersymmetric situations higher loop contributions are
expected.

\subsection{Explicit models}

Here we construct an example with Standard Model gauge group and three
quark-lepton families, in order to illustrate how our more general starting
point overcomes the difficulty found in \cite{bgkl} to obtain three
generations. Notice that the model, just like the examples in \cite{bgkl},
contains tachyons, but we prefer not to list them since they are moduli
dependent (and might even disappear for certain regions in parameter space).

We consider six stacks of D6-branes, $K=6$, with multiplicities and
wrapping numbers given by
\beqa
\begin{array}{cccc}
{\bf N_a} & {\bf (n_a^1,m_a^1)} & {\bf (n_a^2,m_a^2)} &
{\bf (n_a^3,m_a^3)} \\
N_1 = 3 & (1,2) & (1,-1) & (1,-2) \\
N_2 = 2 & (1,1) & (1,-2) & (-1,5) \\
N_3 = 1 & (1,1) & (1,0) & (-1,5) \\
N_4 = 1 & (1,2) & (-1,1) & (1,1) \\
N_5 = 1 & (1,2) & (-1,1) & (2,-7) \\
N_6 = 1 & (1,1) & (3,-4) & (1,-5)
%
\end{array}
\eeqa
This choice satisfies the tadpole conditions. The intersection numbers are
\beqa
\begin{array}{ccccc}
I_{12}=3 & I_{13}=-3 & I_{14}=0 & I_{15}=0 & I_{16}=-3 \\
I_{23}=0 & I_{24}=6  & I_{25}=3 & I_{26}=0 & I_{34}=-6 \\
I_{35}=-3 & I_{36}=0 & I_{45}=0 & I_{46}=6 & I_{56}=3
\end{array}
\eeqa
The spectrum under $U(3)\times U(2)\times U(1)^{4}$ is
\beqa
& 3 (3,2)_{[1,-1,0,0,0,0]} + 3({\ov 3},1)_{[-1,0,1,0,0,0]} + 3({\ov
3},1)_{[-1,0,0,0,0,1]} + & \nonumber \\
& + 6(1,2)_{[0,1,0,-1,0,0]} + 3(1,2)_{[0,1,0,0,-1,0]} +
6(1,1)_{[0,0,-1,1,0,0]} +  & \nonumber \\
& + 3(1,1)_{[0,0,-1,0,1,0]} + 6(1,1)_{[0,0,0,1,0,-1]} +
3(1,1)_{[0,0,0,0,1,-1]} &
\eeqa
where subindices give $U(1)$ charges. Out the six $U(1)$'s the diagonal
linear combination decouples, and two of the remaining are anomalous. A
basis of non-anomalous linear combinations (\ref{lincomb}) is provided by
the coefficient vectors
\beqa
\vec{c}= (1,0,0,0,1,0) \;\; ; \;\; \vec{c}=(0,1,1,0,0,0) \;\; ; \;\;
\vec{c}=(0,1,0,0,0,1)
\eeqa
One can check that the non-anomalous linear
combination
\beqa
Q_Y= -\frac 13 Q_1 - \frac 12 Q_2 - Q_3 - Q_5
\eeqa
can play the role of hypercharge. Indeed, the spectrum, showing only
charges under this $U(1)$, is
\beqa
& SU(3)\times SU(2)\times U(1)_Y & \nonumber \\
& 3(3,2)_{1/6} + 3({\ov 3},1)_{-2/3} + 3({\ov 3},1)_{1/3} + 6(1,2)_{-1/2}
+ 3(1,2)_{1/2} + & \nonumber\\
& + 6(1,1)_{1} + 3(1,1)_{0} + 6(1,1)_{0} + 3(1,1)_{-1} &
\eeqa
giving the chiral fermion content of a three-generation standard model
(up to charges under additional $U(1)$ symmetries).

This example illustrates it is relatively easy to do model building in
this framework. Unfortunately, as explained in the introduction, these
models suffer a hierarchy problem, since they are not supersymmetric, and
it is not possible to lower the string scale by making the six-torus
volume large, since this would give rise to too light KK resonances for
gauge bosons. However, we cannot exclude that further modifications of the
setup improve this aspect. It is conceivable to consider spaces
with a small volume region similar to $\IT^6$, where D6-branes wrap
leading to heavy KK excitation, while the volume of the complete space is
much larger. A simple example can be obtained by surgery, taking
a small $\IT^6$, removing a ball in a region away from the branes, and
gluing a throat connecting it to a large volume manifold. Of course, a
concrete realization of this would require much more careful analysis, and
our comment is just intended for illustration. In any event, the problem
in lowering the string scale is not present in the models of D4- and
D5-branes, to be studied in next sections.

\subsection{Stability and Tachyons}

The lowest lying states in the NS sector of an open string stretched between
intersecting D6-branes are given in (\ref{tachdsix}), along with their
masses. These can be tachyonic or not, depending on the angles between the
D-branes. For instance, for $\vartheta_1\leq \vartheta_2+ \vartheta_3$,
$\vartheta_2\leq \vartheta_3+\vartheta_1$, $\vartheta_3\leq \vartheta_1+
\vartheta_2$, $\vartheta_1+\vartheta_2+\vartheta_3\leq 2$, all states at
the intersections have non-negative mass square. In fact, these are the
conditions for the two intersecting 3-cycles be stable against
recombination into a single smooth 3-cycle \cite{angleth}.

In principle there seems to be no obstruction to the existence of
compactifications on $\IT^6$ with D6-branes wrapped on 3-cycles, such that
every intersection fulfills the above conditions, yielding a
four-dimensional non-supersymmetric chiral theory free of tachyons. Such
configurations would be stable against small perturbations, but, carrying
no net charges, may decay to the vacuum by tunneling through a potential
barrier. Such metastable (rather than absolutely stable) non-BPS
configurations could however lead to perfectly sensible phenomenological
models if their lifetime, exponentially suppressed by the barrier height,
is long enough for cosmological standards.

\subsection{Explicit examples of tachyon elimination}

In this section we construct specific models where all intersections are
tachyon-free for certain regions in the six-torus parameter space, namely
the complex structure of the two-tori.

It turns out that it is easier to build such models if the construction
includes an orientifold projection $\Omega R$ (where $R:z_i\to {\ov z}_i$)
as in \cite{bgkl}. The only differences with respect to our configurations
above is that the angle between the tori axis is projected out, the
D6$_a$-branes wrapped on cycles $(n_a^I,m_a^I)$ must have $\Omega R$
orientifold images (denoted D6$a'$-branes) wrapped on cycles $(n_a^I,m_a^I)$,
and the first tadpole condition in (\ref{tadpole}) becomes $\sum_a N_a\,
n_a^1\,n_a^2\,n_a^3=16$ (not counting images). The potential tachyon
masses are however obtained as above.

The model under consideration is one of the four-dimensional constructions
presented in \cite{bgkl}. The sets of D6-branes are given by
\beqa
\begin{array}{cccc}
N_a \quad & {\bf (n_a^1,m_a^1)}\quad & {\bf (n_a^2,m_a^2)}\quad &
{\bf (n_a^3,m_a^3)} \\
N_1=3 & (1,0) & (1,0) & (1,1) \\
N_2=3 & (1,2) & (1,1) & (1,0) \\
N_3=1 & (1,2) & (1,-2) & (1,0) \\
N_4=1 & (1,0) & (1,0) & (10,1)
\end{array}
\label{modsblum}
\eeqa
plus their $\Omega R$ images.
The main advantage in searching tachyon-free models by using constructions
with an $\Omega R$ orientifold projection, is that, as can be appreciated
in (\ref{modsblum}), it allows all integers $n_a^I$ to be positive. This
simplifies the search for tachyon-free regions, since ensures that taking
large ratios $R_1/R_2$ all angles between branes become small, and states
become less tachyonic. For instance, choosing
\beqa
R^1_2/R^1_1=1  \quad ; \quad R^2_2/R^2_1=3/2 \quad ; \quad R^3_2/R^3_1=2
\eeqa
the masses for the scalars (\ref{tachdsix}) at the different intersections
are
\beqa
\begin{array}{ccccc}
{\bf Intersection} & {\bf \alpha'm_1^2} & {\bf \alpha'm_2^2} & {\bf
\alpha'm_3^2} & {\bf \alpha'm_4^2} \\
12,\; 12',\; 21',\; 1'2' & 0.16 & 0.20 & 0.16 & 0.49 \\
13,\; 13',\; 31',\; 1'3' & 0.20 & 0.15 & 0.20 & 0.45 \\
24,\; 24',\; 42',\; 2'4' & 0.01 & 0.05 & 0.30 & 0.64 \\
34,\; 34',\; 43',\; 3'4' & 0.05 & 0.01 & 0.34 & 0.59
\end{array}
\eeqa
We can see that they are all positive, hence the intersections are free of
tachyons, and the system is stable against recombination of the
corresponding cycles.

Pairs of branes with zero intersection number are parallel in some
two-torus. In this model, in the non-generic case that the branes overlap
in this two-torus, open strings stretched between them would lead to
additional tachyons
\beqa
\begin{array}{ccccc}
{\bf Intersection} & {\bf \alpha'm_1^2} & {\bf \alpha'm_2^2} & {\bf
\alpha'm_3^2} & {\bf \alpha'm_4^2} \\
11'         & 0.35 & 0.35 & -0.35 & 0.65 \\
22' &  -0.04 &  0.04  &  0.67 &  0.33 \\
33'  &  0.05 &   -0.05 &  0.75  &  0.25 \\
44'  & 0.06 &   0.06  &  -0.06 &  0.94 \\
14, \; 1'4' & 0.14 & 0.14 & -0.14 & 0.86 \\
14',\; 41' & 0.21 & 0.21 & -0.21 & 0.79 \\
23,\; 2'3'  & 0.36 & -0.36 & 0.36 & 0.64 \\
23',\; 32' & -0.31 &  0.31 & 0.39 & 0.61
\end{array}
\eeqa
However, these states are not tachyonic if the branes are separated beyond
a critical distance in the corresponding two-torus. It is possible that
higher effects, due to brane interactions (one loop in the open string
channel), induce a non-zero attractive force between such non-intersecting
branes, pushing them to the tachyonic region. In any event, this would be
a higher order effect which might be avoided in more complicated models.
Our point here is that tachyons and intersections, which appear at tree
level and are therefore more dangerous, can be eliminated in some models
by a suitable choice of background geometry.

In principle it is possible that this kind of tachyon-free configurations
exist in models without the orientifold projection, even though a
systematic exploration of parameter space is more difficult. We would like
to conclude by pointing out that, since the main difficulty arising from
satisfying the tadpole conditions, the above ideas may have a much simpler
implementation in other contexts, where such conditions are not relevant.
For instance, one may construct a large class of (meta)stable non-BPS
states in type IIB theory on $\IT^6$, by considering D3-branes wrapped on
3-cycles with tachyon-free intersections.

\subsection{Tachyons and Higgs mechanism}

Even if tachyons are present, we would like to point out a quite different
perspective on them, which is actually applicable to more general examples
(among others, those of D4- and D5-branes in coming sections). As in the
more familiar example of brane-antibrane systems (see e.g.
\cite{revnbps,sft}), condensation of open string tachyons may in certain
situations be interpreted as a Higgs mechanism.
In our present context, the tachyon is charged under the gauge groups on
the intersecting branes, and its condensation reduces the gauge symmetry
to that of the recombined brane. From the spacetime viewpoint, it is
physically clear that the tachyon has a potential with a minimum, at which
the energy of the condensate compensates the difference of tensions
between the final and initial states, and at which the tachyon vev breaks
the initial gauge symmetry. Adapting Sen's ideas \cite{revnbps}, the
intersecting branes with the tachyon condensed to its minimum is {\em
exactly} the final configuration of the recombined brane (stretched along
a minimum volume cycle in its homology class) \footnote{This is
particularly clear in the T-dual picture of D9-branes with magnetic
fluxes, where the above process often amounts to annihilation of
topological defects on the D9-brane world-volume. Some remarks on tachyon
condensation as a Higgs mechanism in this picture have appeared in
\cite{bachas}.}

This idea has an important and interesting caveat, in the interpretation
of the inverse process as un-Higgsing. Basically, the final state does not
keep track of what initial state it came from. Hence if the system is
given energy, it will nucleate not only the W bosons corresponding to the
original initial state, but also W bosons of enhanced symmetries associated
to all other possible initial states in the same energy range. However,
there may be situations where one possible initial state is substantially
lighter than the rest. In this situation, a low energy observer, with a
limited range of available energies, would systematically find a single
pattern of gauge symmetry enhancement. This situation is close enough to a
standard Higgs mechanism, so that tachyons may be interpreted as standard
Higgs fields (at least for processes in the appropriate range of energies),
even for electroweak symmetry breaking. A more detailed understanding of
the tachyon potential and dynamics \cite{sft} would help in determining
if such scenario is indeed viable for electroweak or other phenomenological
Higgs mechanisms. For the moment, we just point out the tantalizing
existence of tachyon fields with the quantum numbers of standard model
Higgs fields in some of the models we have explored (see section 4.3
and\cite{afiru2} for further details).

\section{D4-branes wrapping at angles on $\IT^2\times \IC^2/\IZ_N$}

\subsection{Construction}

As discussed in Section 2, configurations of D4-branes wrapped on 1-cycles
in $\IT^2\times \IC^2$ lead necessarily to non-chiral spectra. In this
section we study a simple modification of this basic framework, which
leads to generically chiral four-dimensional gauge field theories on the
D-brane world-volume.

We consider configurations of D4-branes on $\IT^2\times (\IC^2/\IZ_N)$,
where the D4-branes are distributed in stacks of multiplicity $N_a$,
wrapped along one-cycles $\Pi_a$ defined by wrapping numbers $(n_a,m_a)$,
on $\IT^2$, and sitting at the origin in $\IC^2/\IZ_N$. The models admit a
T-dual description in terms of type IIB D5-branes on $\IT^2\times
(\IC^2/\IZ_N)$, with  non-trivial wrapping numbers and fluxes on $\IT^2$.
We usually phrase our results in the D4-brane picture, but translation to
the D5-brane picture is straightforward, as in the models in Section 3.

The twist $\IZ_N$ is generated by a geometric action $\theta$ with twist
vector given by  $v=\frac{1}N(b_1,b_2,0,0)$, where $b_1=b_2$ mod 2 for the variety
to be spin. The supersymmetric case is recovered when $b_2=-b_1$ mod $N$,
hence with twist $v=\frac{1}N(b_1,-b_1,0,0)$. In this case, since $b_1$
and $N$ must be coprime for a $\IZ_N$ action, the orbifold group can be
equivalently generated by the twist $\theta^k$, with $kb_1=1$ mod $N$,
which has the more familiar twist vector $v=\frac{1}N(1,-1,0,0)$.

We would like to emphasize that we imagine this framework as a local
description of the configuration near the location of the branes.
Globally, the local configuration above may be embedded in a spacetime of
the form $\IT^2\times {\bf B}$, where ${\bf B}$ is a four-dimensional
space (not necessarily Calabi-Yau) with a $\IC^2/\IZ_N$ singularity at
which the D-branes sit. More generally, the complete space may not be
globally a product, but rather a torus bundle over ${\bf B}$, or even a
torus fibration, as long as singular fibers are away from the D-brane
location. Our configuration is a good local description in these cases,
and completely controls the structure of the D-brane world-volume gauge
theory.

Let us briefly mention another interesting aspect. These configurations
admit a seemingly simple lift to M-theory, as a set of M-theory fivebranes
sitting at a $\IC^2/\IZ_N$ singularity, and wrapped on a two-cycle in
$\IT^2\times \IS^1$. Obviously, a detailed description of the model in
M-theory will involve a number of interesting subtleties, on which our
analysis may shed some light. Note that the existence of this
six-dimensional parent theory, which reduces to the four-dimensional field
theory after compactification, is not in contradiction with chirality in
the latter. The higher dimensional theory is not a conventional field
theory, and in fact four dimensional chiral states arise from membranes
stretched between M5-branes and wrapped on $\IS^1$, i.e. they do not
descend by KK reduction from any six-dimensional field.

Let us describe the computation of the spectrum in our configuration. The
closed string sector is computed using standard orbifold techniques. In the
supersymmetric case, it gives rise to an $D=4$ $\NN=4$ $U(1)^{N-1}$ gauge
multiplet. In the non-supersymmetric case, the main feature is that it
leads to tachyons in the NS-NS sector. Their interpretation is, as usual
with closed string tachyons, not understood, and we will have nothing new
to say about them. Nevertheless, we choose to study these models and in
particular their open string spectrum even for non-supersymmetric
singularities.

The $\IZ_N$ action may be embedded in the $U(N_a)$ gauge degrees of
freedom of the $a^{th}$ stack of D4-branes, through a unitary matrix of
the form
\beq
\gamma_{\theta,a} \ = \ \diag(\id_{N_a^0},e^{2\pi i\frac 1N} \id_{N_a^1},
\ldots ,e^{2\pi i\frac {N-1}N} \id_{N_a^{N-1}})
\label{chanpaton}
\eeq
with $\sum_i N_a^i=N_a$.

Let us compute the spectrum in the open string sector. In the $4_a4_a$
sector, the massless states surviving the GSO projection, along
with their behaviour under the $\IZ_N$ twist, are
\beqa
\begin{array}{cccc}
{\rm\bf NS}\; {\rm\bf State} & \IZ_N\; {\rm\bf phase} &\quad
{\rm\bf R}\; {\rm\bf State} & \IZ_N\; {\rm\bf phase}  \\
(\pm 1,0,0,0) & e^{\pm 2\pi i\frac{b_1}{N}} &
\pm \frac 12(-,+,+,+) & e^{\mp \pi i\frac{b_1-b_2}N} \\
(0,\pm 1,0,0) & e^{\mp 2\pi i\frac {b_2}{N}} &
\pm \frac 12(+,-,+,+) & e^{\pm \pi i\frac{b_1-b_2}N} \\
(0,0,\pm 1,0) & 1         &
\pm \frac 12(+,+,-,+) & e^{\pm \pi i\frac{b_1+b_2}N} \\
(0,0,0,\pm 1) & 1        &
\pm \frac 12(+,+,+,-) & e^{\pm \pi i\frac{b_1+b_2}N} \\
\end{array}
\eeqa
The open string spectrum is obtained by keeping states invariant under the
combined geometric plus Chan-Paton $\IZ_N$ action \cite{dm}. After the
$\IZ_N$ projections, the resulting gauge group and matter fields are
\beqa
{\rm\bf Gauge\; Bosons} & \quad \prod_{a=1}^K \prod_{i=1}^N U(N_a^i)
\nonumber\\
{\rm\bf Cmplx.\; Scalars} & \quad \sum_{a=1}^K \sum_{i=1}^N \;
[\, (N_a^i,{\ov N}_a^{i+b_1}) + (N_a^i,{\ov N}_a^{i+b_2}) \, ] \nonumber\\
{\rm\bf  Left \; Fermion} & \quad \sum_{a=1}^K \sum_{i=1}^N \; [\;
(N_a^i,{\ov N}_a^{i+(b_1-b_2)/2}) + (N_a^i,{\ov N}_a^{i-(b_1-b_2)/2})
\nonumber \\
{\rm\bf Right\; Fermion} & \quad \sum_{a=1}^K \sum_{i=1}^N \; [\;
(N_a^i,{\ov N}_a^{i+(b_1+b_2)/2}) + (N_a^i,{\ov N}_a^{i-(b_1+b_2)/2})
\eeqa
Notice that this piece of the spectrum is generically non-supersymmetric,
and always non-chiral. In the supersymmetric case, $v=(1,-1,0,0)/N$, this
sector preserves $\NN=2$ supersymmetry in the four-dimensional field
theory. The above fields form the multiplets
\beqa
{\bf \NN=2}\; {\rm\bf Vector} &\quad \prod_{a=1}^K \prod_{i=1}^N U(N_a^i)
\nonumber\\
{\bf \NN=2}\; {\rm\bf Hyper} & \sum_{a=1}^K \sum_{i=1}^N (N_a^i,{\ov
N}_a^{i+1})
\eeqa
In the $4_a4_b$ sector, open strings are twisted by the angle formed by
the branes, denoted $\vartheta$, resulting in a sector twisted by the
shift $(\vartheta,0,0,0)$. Assuming $0 \leq \vartheta\leq 1$, tachyonic
and massless states, along with their $\IZ_N$ phases, are
\beqa
\begin{array}{ccc}
{\rm\bf Sector} & {\rm\bf State} & \IZ_N\; {phase} \\
{\rm NS} & (-1+\vartheta,0,0,0) & 1 \\
{\rm R}  & (-\s2+\vartheta,-\s2,-\s2,+\s2) &
e^{-2\pi i\frac{(b_1+b_2)}{2N}} \\
         & (-\s2+\vartheta,-\s2,+\s2,-\s2) &
e^{-2\pi i\frac{(b_1-b_2)}{2N}}\\
         & (-\s2+\vartheta,+\s2,-\s2,-\s2) & e^{2\pi i\frac{(b_1-b_2)}{2N}}\\
         & (-\s2+\vartheta,+\s2,+\s2,+\s2) & e^{2\pi i\frac{(b_1+b_2)}{2N}}
\end{array}
\eeqa
This piece of the spectrum is non-supersymmetric, even for supersymmetric
$\IZ_N$ twists. The NS states are tachyonic, with $\alpha' M^2$ equal to
$-\frac 12 |\vartheta|$, and signal an instability against recombining
intersecting D4-branes with same Chan-Paton eigenvalue. Hence, they may be
avoided by suitable choices of the $\IZ_N$ actions $\gamma_{\theta, 4_a}$.
A different possibility is to interpret these tachyons as triggering
breaking of gauge symmetries by a Higgs mechanism, as mentioned in section
3.4. The R states are massless, and provide a set of chiral fermions in
the model. Notice that the antiparticles of these states appear in the
$4_b 4_a$ sector, which is twisted by $(-\vartheta,0,0,0)$.

In these sectors the spectrum generically appears in several replicas,
whose number is given by the intersection number $I_{ab}$ of the one-cycles
$\Pi_a$ and $\Pi_b$ in $\IT^2$,
\beqa
I_{ab}=n_am_b-m_an_b
\eeqa
The spectrum after the Chan-Paton projections is given by
\beqa
{\rm\bf Cmplx.\; Tachyons} & \quad \sum_{a<b} \sum_{i=1}^N \; I_{ab}\times
(N_a^i,{\ov N}_b^i) \nonumber \\
{\rm\bf Left\; Fermion} & \quad \sum_{a<b} \sum_{i=1}^N \;  I_{ab}\times\;
[\; (N_a^i,{\ov  N}_b^{i+(b_1+b_2)/2} + (N_a^i,{\ov N}_b^{i-(b_1+b_2)/2}
\; ] \nonumber \\
{\rm\bf Right\; Fermion} & \quad \sum_{a<b} \sum_{i=1}^N \;I_{ab}\times
[\; (N_a^i,{\ov  N}_b^{i+(b_1-b_2)/2} + (N_a^i,{\ov  N}_b^{i-(b_1-b_2)/2}
\; ]
\label{chispfour}
\eeqa
which is non-supersymmetric and generically chiral. Therefore the resulting
field theories may lead to phenomenologically interesting models. In fact,
in Section 4.3 we construct an explicit example with Standard Model group
and three quark lepton generations.

\subsection{Tadpoles and anomalies}

\subsubsection{Tadpole cancellation conditions}

The consistency conditions (RR tadpole cancellation conditions) for these
configurations are easily computed, and read
\beqa
\begin{array}{c}
\prod_{r=1,2} \sin(\pi kb_r /N) \sum_{a=1}^K n_a  \Tr
\gamma_{\theta^k,4_a} = 0 \nonumber \\
\prod_{r=1,2} \sin(\pi k b_r/N) \sum_{a=1}^K m_a  \Tr
\gamma_{\theta^k,4_a} = 0
\end{array}
{\rm for} \; k=1,\ldots, N-1
\label{tadpofour}
\eeqa
There is no constraint associated to $k=0$, since the untwisted tadpole is
associated to a flux that can escape along the non-compact dimensions of
$\IC^2/\IZ_N$.

These conditions can be interpreted geometrically, at least for
supersymmetric singularities, by regarding the fractional \cite{fract}
D4$_a^s$-branes (i.e. the set of D4$_a$ branes associated to the phase
$e^{2\pi i\frac sN}$ in $\gamma_{\theta,4_a}$) as D6-branes wrapped on
the 1-cycle $[\Pi_a]=n_a[a]+m_a[b]$ in $\IT^2$ times the $s^{th}$ collapsed
two-cycle $[\Sigma_s]$ in the singularity. The conditions above amount to
the vanishing of the total homology class
\beqa
\sum_{a=1}^K \sum_{s=0}^{N-1} N_a^s\, [\Pi_a]\otimes [\Sigma_s] = 0
\eeqa
Since $\sum_{s=0}^{N-1} [\Sigma_s]=0$, one can increase the $N_a^s$ by an
$s$-independent (but possibly $a$-dependent) amount and still satisfy the
homological condition. Hence, the Chan-Paton matrices for $k=0$ are
unconstrained. Note that regarding branes at singularities as branes
wrapped on collapsed cycles, our models of D4-branes become a degenerate
version of D6-branes wrapped on 3-cycles in a curved ambient space, and
following footnote 4, our results here are reminiscent of those in section
3.2.

\subsubsection{Anomaly cancellation}

The spectrum of the model is generically chiral, and has potential gauge
anomalies. In analogy with the case of D6-branes on $\IT^6$, cancellation
of the anomalies due to chiral fermions at each intersection would be achieved
by an inflow mechanism. Since the intersections sit at the singularity in
transverse space, this inflow mechanism would be more involved and
interesting, but still tractable. Leaving aside its study, we prefer to
center on the compactified effective four-dimensional description of
anomaly cancellation.

The cancellation of cubic non-abelian anomalies for $SU(N_a^i)$
gives the conditions
\beqa
\sum_{b=1}^K I_{ab} (-N_b^{i+(b_1+b_2)/2} -N_b^{i-(b_1+b_2)/2}
+N_b^{i+(b_1-b_2)/2} +N_b^{i-(b_1-b_2)/2}) = 0
\eeqa
These conditions should follow from the tadpole cancellation conditions.
In fact, using (\ref{chanpaton}) we can rewrite \beqa N_b^i = {1\over N}
\sum_{k=0}^{N-1} e^{-2\pi i \frac{ki}{N}} \Tr\gamma_{\theta^k,4_b}
\label{fourier}
\eeqa
as in \cite{leroz}, and the anomaly cancellation conditions read
\beqa
\prod_{r=1,2 } \sin(\pi kb_r/N) \sum_{b=1}^K I_{ab} \Tr
\gamma_{\theta^k,4_b} = 0
\label{anomfour}
\eeqa
These conditions are indeed guaranteed by the tadpole conditions
(\ref{tadpofour}), but, as usual, are much milder than the latter.

Let us consider cancellation of mixed $U(1)$ anomalies, which involves
a generalized Green-Schwarz mechanism,
mediated by $2(N-1)$ fields, corresponding to the integration of $N-1$
twisted RR-fields along the two independent 1-cycles in the $\IT^2$.

In fact, one can compute the mixed anomaly between the $U(1)_{ai}$ and
$SU(N_b^j)$ using the chiral piece of the spectrum (\ref{chispfour}).
After removing a vanishing piece proportional to the non-abelian anomaly,
there remains
\beqa
{\cal A}_{ai,bj} = \frac 12\, N_a^i\, I_{ab}\,
(\delta_{j,i+(b_1+b_2)/2}+\delta_{j,i-(b_1+b_2)/2}
-\delta_{j,i+(b_1-b_2)/2}-\delta_{j,i-(b_1-b_2)/2})
\label{mixedfour}
\eeqa
Substituting the discrete Fourier transform representation of the
Kronecker deltas, as in \cite{iru}, the anomaly acquires the nice
factorized form
\beqa
A_{ai,bj} = i N_a^i I_{ab} \frac 1N \sum_{k=1}^{N-1}
4\prod_{r=1,2}\sin(\pi kb_r/N)\; e^{2\pi i\frac{ki}{N}} \;
 e^{-2\pi i\frac{kj}{N}}
\label{factori}
\eeqa
The anomaly may therefore be cancelled by exchange of the four dimensional
fields obtained by integrating  over the two one-cycles
in $\IT^2$ the RR twisted forms, which give the four-dimensional couplings
\beqa
& c_k\, N_a^i\, n_a \int_{M_4} \Tr (\gamma_{\theta^k,4_a}\lambda_i)\,
C_2^{(k)}\wedge \tr F_{a,i} \; & ;\;\;
c_k \, m_b \int_{M_4} \Tr (\gamma_{\theta^k,4_b} \lambda_j^2)\,
C_0^{(k)}\wedge \Tr F_{b,j}^2 \nonumber\\
& - c_k \, N_a^i\, m_a \int_{M_4} \Tr (\gamma_{\theta^k,4_a}\lambda_i)\,
B_2^{(k)}\wedge \tr F_{a,i} \quad & ;\quad
c_k\, n_b \int_{M_4} \Tr (\gamma_{\theta^k,4_b} \lambda_j^2)\,
B_0^{(k)}\wedge \Tr F_{b,j}^2 \nonumber
\eeqa
where $\lambda$ denotes the CP wavefunction of the gauge boson state. The
prefactors $c_k=(\prod_r \sin(\pi k b_r/N))^{1/2}$ can be thought of as
arising from ${\hat A}^{1/2}$ in D-brane couplings \cite{lidouglas}, and
have been explicitly computed in string theory in e.g. \cite{abd,scruse}.
Since $B_2$ and $B_0$, and $C_2$ and $C_0$ are Hodge
dual in four dimensions, the sum over GS diagrams has the structure
(\ref{factori}). The GS mechanism is analogous to that for D6-branes on
$\IT^6$, as is manifest from the appearance of the intersection number,
the main difference being that the exchanged fields belong to twisted
sectors of the $\IC^2/\IZ_N$ factor.

The above results can be interpreted geometrically by regarding the
fractional D4$_a$-branes as D6-branes wrapped on collapsed cycles $\Sigma_s$
of the singularity and the one-cycle $\Pi_a$ in $\IT^2$. This is simplest
in the more familiar supersymmetric case where the anomaly is given by
\beqa
A_{ai,bj} = \frac 12 N_b^j I_{ab}
(2\delta_{j,i}-\delta_{j,i+1}-\delta_{j,i-1})
\label{mixedfoursusy}
\eeqa
The collapsed two-cycles have intersections given by (minus) the Cartan
matrix of the (affine) $\hat{A}_{N-1}$ algebra, $C_{ij}=[\Sigma_i]\cdot
[\Sigma_j] = -2\delta_{ji}+ \delta_{j,i+1} +\delta_{j,i-1}$. Hence, the
intersection number of D6-branes wrapped on cycles $[\Pi_a]\otimes[\Sigma_i]$
and $[\Pi_b]\otimes[\Sigma_j]$ is $I_{ab}C_{ij}$. Introducing a composite
index $I$ grouping together indices $a$ and $i$, we can express the mixed
anomaly (\ref{mixedfoursusy}) as
\beqa
A_{IJ} = \frac 12 N_I {\cal I}_{IJ}
\eeqa
where ${\cal I}_{IJ}$ denotes the 3-cycle intersection form. The situation is
hence analogous to that in section 3.2. As suggested in footnote 4, the GS
cancellation mechanism in section 3.2 can be directly
translated, with the obvious modifications, reproducing the cancellation
of anomalies in the present context. Since here the wrapped 3-cycles are
exceptional divisors of the singularity, the forms mediating the GS
mechanism arise as twisted states in string theory. The above geometric
interpretation follows also for non-supersymmetric $\IZ_N$ twist, by using
the corresponding intersection matrix, obtained from (\ref{mixedfour}).

As in section 3.2, anomalous U(1)'s get a mass of the order of the string
scale. To find non-anomalous $U(1)$'s, we consider linear combinations
\begin{equation}
Q= \sum _{a=1}^K \sum _{j=0}^{N-1} \frac{c_{a,j}}{{N_a^j}} Q_{a,j}
\end{equation}
(we choose $c_{b,i}=0$ if the corresponding group is not present, namely
if $N_a^i=0$). Non-anomalous $U(1)$s can again be found as zero modes of the
(generalized) intersection matrix. In fact, we can be slightly more
explicit. Taking the supersymmetric singularity case for concreteness, and
using the expression (\ref{factori}), anomaly-free linear combinations
satisfy
\beqa
\frac 1N \sum_{k=1}^{N-1} e^{-2\pi i\frac{kj}{N}}\, \sin^2(\pi k/N)\;
\sum_{a=1}^K I_{ab} \sum_{i=0}^{N-1} e^{2\pi i\frac{ki}{N}} \; c_{a,i} = 0
\eeqa
for each $b=1,\dots K$ and $j=0,\dots,N-1$. A useful trick is to
perform the change of coordinates \cite{aiqu} $r_{a,k}=\sum_{i=0}^{N-1}
e^{2\pi i\frac{ki}{N}}c_{a,i}$, and obtain the conditions
\begin{equation}
\label{afree2}
\sin^2 (\pi k/N) \sum_{a=1}^K\, I_{ab}\, r_{a,k} = 0
\end{equation}
A set of solutions is given by choosing, for a fixed $a$,
$r_{a,k}=\delta_{k,0}$, and $r_{b,k}=0$ for $b\neq a$. The resulting
generator is
\begin{equation}
Q_a= \sum_{i=0}^{N-1} \frac{Q_{a,i}}{N_a^i}
\end{equation}
Another combination is obtained by choosing $c_{a,i}= N_a^i$, or equivalently
by $r_{a,k}= \Tr \gamma_{\theta^k,4_a}$, namely
\begin{equation}
Q= \sum_{a=1}^K \sum_{i=0}^{N-1} Q_{a,i}
\end{equation}
Depending on the details of the orbifold group, there may be additional
non-anomalous U(1)'s. These are most easily determined by directly
computing the zero modes of the anomaly matrix in each case.

\subsection{Explicit models}

In the present context it is not possible to get rid of all the tachyons
while maintaining a chiral fermion spectrum. A general argument goes as
follows. Since tachyons arise in $4_a4_b+4_b4_a$ sectors from strings
stretching between D4-branes with same Chan-Paton phase, to avoid tachyons
we must consider models where any two intersecting branes have no common
Chan-Paton eigenvalues. Consider
models with $N$ stacks of D4$_a$-branes, hence $K=N$, at a $\IC^2/\IZ_N$
singularity, with twist e.g. $v=(1,-1)/N$, and wrapped on arbitrary
1-cycles $\Pi_a$ on $\IT^2$, and choose the Chan-Paton
embeddings
\beqa
\gamma_{\theta,4_a} = e^{2\pi i\frac aN} \id_{N_a}
\eeqa
hence $N_a^i=N_a \delta_{a\,i}$ (more general choices can be treated
analogously). The spectrum one naively obtains seems chiral, but the
tadpole conditions
\beqa
\sum_{a=1}^N e^{2\pi i \frac{ka}{N}} N_a [\Pi_a] = 0 \ {\rm for}\;
k=1,\ldots,
N-1
\eeqa
turn out to be very constraining. By discrete Fourier transforming, they
imply that all $[\Pi_a]$ are actually identical, so all D4-branes are
parallel, leading to non-chiral spectra.

Allowing a non-supersymmetric singularity may relax the tadpole conditions,
but introduces (closed string) tadpoles. Also, the case $K<N$ reduces to
the above with some $N_a=0$, while additional branes ($K>N$) necessarily
repeat eigenvalues and must be non-intersecting, i.e. parallel, to the
existing ones to avoid tachyons.

Allowing for some tachyons in the model, however, one can obtain large
classes of models with chiral spectrum, which moreover can be quite close
to realistic models. Let us discuss a simple explicit model, which
illustrates a possible model building strategy.

Since $\IZ_2$ leads to vector-like models, let us consider sets of D4-branes
at $\IT^2\times \IC^2/\IZ_3$, with twist $v=(1,-1,0,0)/3$. A typical
tachyon-free and hence non-chiral model would have three stacks of D4-branes,
with $\gamma_{\theta,4_a}=e^{2\pi i\, a/3}\id$, and parallel cycles.
Consider e.g. the stack with $\gamma_{\theta}=\id$ with multiplicity $3$
and cycle $(1,0)$, and the remaining two with multiplicity $1$ and cycle
$(3,0)$, yielding a gauge group $U(3)\times U(1)\times U(1)$ with
vector-like matter. To get chirality, we must allow one of these sets to
split into several intersecting stacks, a process which also implies the
appearance of tachyons (which would trigger the recombination to the
original configuration). Let us build this enlarged model with Standard Model
gauge group, namely including a stack with group $U(2)$, and with
triplicated intersections. A possible choice is to split the brane
with $\gamma_{\theta}=e^{2\pi i/3}$ wrapped on $(3,0)$, into two branes
wrapped on $(1,3)$, one brane on $(0,-3)$ and one brane on $(1,-3)$. Hence
we end up with
\beqa
\begin{array}{ccc}
{\rm\bf Multiplicity} & {\rm\bf Cycle} & {\rm\bf CP\; phase} \\
N_1=3 & (1,0) & 1 \\
N_2=2 & (1,3) & e^{2\pi i/3} \\
N_3=1 & (0,-3) & e^{2\pi i/3} \\
N_4=1 & (1,-3) & e^{2\pi i/3} \\
N_5=1 & (3,0) &  e^{2\pi i\, 2/3}
\end{array}
\eeqa
The chiral spectrum contains left-handed fermions transforming under
$U(3)\times U(2)\times U(1)_3^3\times U(1)_4\times U(1)_5^3$ as
\beqa
&& 3(3,2)_{[1,-1,(0^3),0,(0^3)]} +
(\bar 3,1)_{[-1,0,{\underline{1,0,0}},0,(0^3)]} +
3(\bar 3,1)_{[-1,0,(0^3),1,(0^3)]}+\nonumber \\
& & 2(1,2)_{[0,1,{\underline{-1,0,0}},0,(0^3)]}
+12(1,2)_{[0,1,(0^3),-1,(0^3)]} +
3(1,2)_{[0,-1,(0^3),0,{\underline{1,0,0}}]} \nonumber \\
& & + 2(1,1)_{[0,0,{\underline{-1,0,0}},1,(0^3)]} +
(1,1)_{[0,0,{\underline{1,0,0}},0,{\underline{-1,0,0}}]}
+ 3(1,1)_{[0,0,(0^3),1,{\underline{-1,0,0}}]}
\eeqa
where underlining means permutation.
Besides the diagonal combination, which decouples, there are six
non-anomalous $U(1)$ linear combinations. One of them, given by
\beqa
Q_{Y}\ & =&\ -{1\over 3}Q_1 -{1\over 2}Q_2 - Q_4-(Q_5^{(1)} +Q_5^{(2)} +
Q_5^{(3)})
\label{bhyper}
\eeqa
provides correct hypercharge assignments for the above theory, which
therefore has the chiral content of a three-generation standard model.
Indeed, highlighting the charges under this $U(1)$, the fermion spectrum
is
\beqa
& 3(3,2)_{1/6} + 3(\bar 3,1)_{1/3} + 3(\bar 3,1)_{-2/3}+
15(1,2)_{-1/2} + & \nonumber \\
& 12(1,2)_{1/2} + 6(1,1)_{-1} +9(1,1)_{1} + 9(1,1)_{0} &
\eeqa

Further properties of these models will be discussed in
\cite{afiru2}. Here let us simply point out that, in models constructed
using the above strategy, the tachyons trigger the recombination of branes
involving the $U(2)$ factor, and therefore have the gauge quantum numbers
of standard model Higgs fields. These models therefore illustrate that
tachyonic modes may be phenomenologically interesting (to trigger
electroweak or other extended symmetry breakings), and that in this
class of models they are linked to the existence of chiral fermions.

As mentioned in the introduction, even though the models are
non-supersymmetric, the hierarchy problem is avoided by considering a low
string scale and a compactification with large volume for the space
transverse to the two-torus.

\section{D5-branes wrapping at angles on $ (\IT^2)^2\times \IC/\IZ_N$}

\subsection{Construction}

For completeness, in this section we center on a last type of configuration,
similar to those in the preceding section, and also leading to
four-dimensional chiral theories. We consider configurations of D5-branes
in $\IT^4\times (\IC/\IZ_N)$, where the D5-branes sit at the origin in
$\IC/\IZ_N$, and are grouped in stacks of multiplicity $N_a$ wrapped on
2-cycles defined by $(n_a^I,m_a^I)$, with $I=1,2$, in a factorizable
$\IT^4$.

The $\IZ_N$ action on the third dimension is encoded in the twist vector
of the form $v=\frac 1N (0,0,2,0)$ for the variety to be spin. The closed
string sector necessarily contains tachyons in its twisted sector, whose
interpretation is unclear. Nevertheless, we proceed studying these models.
The Chan-Paton twist matrices have the general form
\beq
\gamma_{\theta,a} \ = \ \diag(\id_{N_a^0},e^{2\pi i\frac 1N} \id_{N_a^1},
\ldots ,e^{2\pi i\frac {N-1}N} \id_{N_a^{N-1}})
\eeq
with $\sum_i N_a^i=N_a$. The lowest lying states in the $5_a5_a$ open
string NS and R sectors, along with their $\IZ_N$ phases, are
\beqa
\begin{array}{ccccc}
{\rm\bf NS}\; {\rm\bf State} & \IZ_N\; {\rm\bf phase} & \quad ;\quad &
{\rm\bf R}\; {\rm\bf State} & \IZ_N\; {\rm \bf phase} \\
(\pm 1,0,0,0) & 1 & &
\pm \frac 12(-,+,+,+) & e^{\pm 2\pi i\frac 1N} \\
(0,\pm 1,0,0) & 1 & &
\pm \frac 12(+,-,+,+) & e^{\pm 2\pi i\frac 1N} \\
(0,0,\pm 1,0) &   e^{\pm 4\pi i\frac 1N} & &
\pm \frac 12(+,+,-,+) &  e^{\mp 2\pi i\frac 1N} \\
(0,0,0,\pm 1) & 1  & &
\pm \frac 12(+,+,+,-) &  e^{\pm 2\pi i\frac 1N}  \\
\end{array}
\eeqa
The spectrum is non-supersymmetric. The fourth NS state leads to $
\prod_{a=1}^K \prod_{i=1}^N U(N_a^i)$ gauge bosons, while the remaining
give a set of scalars in bifundamental or adjoint representations.
In the R sector, no state is invariant under $\IZ_N$, and the model
contains no gauginos. On the other hand, it contain a non-chiral
set of fermions in diverse bifundamental representations. Summarizing, the
spectrum contains the following fields
\beqa
{\rm\bf Gauge\; Bosons} &\quad \prod_{a=1}^K \prod_{i=1}^N U(N_a^i)
\nonumber\\
{\rm\bf Real\; Scalars } & \sum_{a=1}^K \sum_{i=1}^N
[\; (N_a^{i},{\ov N}_a^{i+2}) + 2\times  {\rm Adj}_{a,i}\; ] \nonumber\\
{\rm\bf Right\; Fermion} & \sum_{a=1}^K \sum_{i=1}^N (N_a^{i},{\ov
N}_a^{i+1}) \nonumber\\
{\rm\bf Left\;Fermion} & \sum_{a=1}^K \sum_{i=1}^N ({\ov N}_a^{i+1},N_a^{i})
\label{D5abspec}
\eeqa

In the $5_a5_b$ sector, open strings are twisted by the angle formed by
the branes in the two-tori, encoded in a twist vector $(\vartheta_1,
\vartheta_2,0,0)$. The lowest lying states, along with their behaviour
under $\IZ_N$, are (assuming $0\leq\vartheta_I\leq 1$)
\beqa
\begin{array}{ccc}
{\rm\bf Sector} & {\rm\bf State} & \IZ_N  {\rm\bf phase} \\
{\rm NS} & (-1+\vartheta_1,0,0,0) & 1 \nonumber \\
         & (0,-1+\vartheta_2,0,0) & 1 \nonumber \\
{\rm R}  & (-\s2+\vartheta_1,-\s2+\vartheta_2,+\s2, -\s2) &
e^{2\pi i\frac 1N} \\
         &  (-\s2+\vartheta_1,-\s2+\vartheta_2,-\s2, +\s2) &
e^{-2\pi i\frac 1N}
\end{array}
\eeqa
Recall that at most one of the two NS states is tachyonic (both are
massless for $|\vartheta_1|=|\vartheta_2|$), while fermions are massless.

The spectrum of tachyonic and massless states, after the Chan-Paton
projections, and taking into account the multiplicity due to the
intersection numbers
\beqa
I_{ab} =I_{ab}^1 I_{ab}^2= (n_a^1 m_b^1- m_a^1\ n_b^2)
( n_a^2 m_b^2- m_a^2 n_b^2)
\label{interfive}
\eeqa
is given by
\beqa
{\rm\bf Cmplx.\; Tachyons} & \quad \sum_{a<b} \sum_{i=1}^N \; I_{ab}\times
(N_a^i,{\ov N}_b^i) \nonumber \\
{\rm\bf Left\; Fermion} & \quad \sum_{a<b} \sum_{i=1}^N \;
I_{ab}\times(N_a^i,{\ov  N}_b^{i+1}) \nonumber \\
{\rm\bf Right\; Fermion} & \quad \sum_{a<b} \sum_{i=1}^N \;  I_{ab}\times
(N_a^i,{\ov  N}_b^{i-1})
\eeqa
In the case $ |\vartheta_2|=|\vartheta_1|$ we would have two bosonic
massless states instead of the above tachyon.

\subsection{Tadpoles and anomalies}

The analysis of tadpole and anomaly cancellation is similar to that for
configurations in section 4.2, hence our discussion is more sketchy.

Tadpole cancellation conditions read
\beqa
& \sin(4\pi k/N)\, n_a^1\, n_a^2\,  \Tr \gamma_{\theta^k,4_b} = 0 \quad ;
& \quad \sin(4\pi k/N)\, m_a^1\, n_a^2\, \Tr \gamma_{\theta^k,4_b} = 0
\nonumber\\
& \sin(4\pi k/N)\, n_a^1\, m_a^2\, \Tr \gamma_{\theta^k,4_b} = 0 \quad ;
& \sin(4\pi k/N)\, m_a^1\, ma^2\, \Tr \gamma_{\theta^k,4_b} = 0
\eeqa
and clearly have the interpretation of cancellation of charges analogous
to that for equations (\ref{tadpofour}).

These conditions must ensure the consistency of the low-energy
four-dimensional field theory on the D-brane world-volume. In particular,
the cancellation of cubic non-abelian chiral anomalies for $SU(N_a^i)$
reads
\beqa
\sum_{b=1}^K I_{ab} ( N_b^{i+1} - N_b^{i-1}) =0
\label{nonabelianD5}
\eeqa
or, equivalently, by performing the discrete Fourier transform
(\ref{fourier}),
\begin{equation}
\sin(4\pi k/N) I_{ab}  \Tr \gamma_{\theta^k,5_b} = 0
\end{equation}
By substituting (\ref{interfive}) in this equation, we see it is implied
by the tadpole constraints.

Using the spectrum (\ref{D5abspec}) it is easy to compute the mixed
anomalies between $U(1)_{ai}$ and $SU(N_{bj})$. We obtain
\beqa
A_{ai,bj} = \frac 12\, N_a^j\, I_{ab}
(\delta_{j,i+1}-\delta_{j,i-1})= i\, N_a^i\, I_{ab}\, \frac 1N
\sum_{k=1}^{N-1}
\, \sin(2\pi k/N) \, e^{2\pi i\frac{ki}{N}} \, e^{-2\pi i\frac{kj}{N}}
\label{D5mixed}
\eeqa
where, again, the second equality shows the residual anomaly has a
factorized structure, which can be cancelled by a GS mechanism mediated by
four-dimensional fields obtained by integrating twisted RR fields
on diverse two-cycles in $\IT^4$.

The existence and form of non-anomalous (and therefore massless) $U(1)$
linear combinations can be carried out in complete analogy with that in
section 4.2.2

\section{Final comments and outlook}

In this paper we have studied the construction of four-dimensional chiral
string compactifications with gauge sector localized on D-branes wrapped
on non-trivial cycles in the internal space. Specifically, we have studied
configurations of D$(3+n)$-branes wrapped on $n$-cycles in $\IT^{2n}\times
\IC^{3-n}/\IZ_N$, where the last factor should be understood as a local
model of a singularity within a compact $(6-2n)$-dimensional variety,
so that correct four-dimensional gravity is recovered. Several properties
(like the anomaly cancellation mechanisms) however hold in more general
setups.

The configurations allow a bottom-up approach to embedding realistic gauge
sectors in string theory models, in the sense explained in \cite{aiqu}. In
fact, the configurations are a natural extension of the work on D3-branes
at threefold singularities (e.g. $\IC^3/\IZ_N$) in \cite{aiqu}. However,
we have found a number of interesting differences, and original features
in the configurations considered in this paper.

Our results in this paper extend the early results in \cite{bdl} on
intersecting branes to the context of compact models, leading to a large
class of non-supersymmetric chiral four-dimensional models. We have
provided a simple set of rules to construct explicit models, and studied
there general features. One amusing feature is that, as observed in
\cite{bgkl}, compact models of intersecting branes lead naturally to
replication of the chiral fermion content, due to the multiple
intersections between different
wrapped branes. In fact, we have used this property to construct explicit
three generation models with realistic gauge groups.

In analogy with other string compactifications, we have found a rich
structure of mixed $U(1)$ anomalies. We have shown that they are cancelled
by a generalized GS mechanism mediated by untwisted or twisted RR fields.
While the GS mediation by the latter is familiar from type IIB orientifolds
\cite{sagnan,iru}, the former (valid for D6-brane models) is rather
unusual and interesting. We expect it have relevant phenomenological
applications.

Finally, we have discussed that although the models are non-supersymmetric,
they can be used for phenomenological purposes without a hierarchy problem,
by simply lowering the string scale, and enlarging the volume transverse
to the D-branes. Lack of supersymmetry also induces the appearance of
tachyons, for which we have suggested elimination mechanism, and a
tantalizing phenomenological application in certain regimes.

Leaving further phenomenological properties of these configurations
for discussion in \cite{afiru2}, we conclude hoping these results are
helpful in the construction of new open string vacua, and in their
phenomenological application in the brane-world scenario.

\bigskip

\centerline{\bf Acknowledgements}

We are grateful to R.~Blumenhagen, B.~K\"ors, D.~L\"ust, F.~Marchesano,
J.~Mas, and F.~Quevedo for useful discussions. A.~M.~U. thanks the Instituto
Balseiro, CNEA, Centro At\'omico Bariloche, Argentina, and the Dpto
F\'{\i}sica Te\'orica, Universidad Aut\'onoma de Madrid, for hospitality,
and M.~Gonz\'alez for encouragement and support. Work by G. A.  is
partially supported by ANPCyT grant 03-03403. L.E.I. and R.R. are
partially supported by CICYT (Spain) and  the European Commission (grant
ERBFMRX-CT96-0045).

\end{document}